\documentclass[pra,aps,showpacs,twocolumn,tightenlines,epsfig]{revtex4}
\usepackage{graphicx}
\usepackage{amsmath}
\begin{document}
\title{Generation of a superposition of
multiple mesoscopic states of radiation in a resonant cavity}
\author{P. K. Pathak~$^*$ and G. S. Agarwal \footnote{On leave:
Physical Research Laboratory, Navrangpura, Ahmedabad-380 009,
India}}
\address{Department of Physics, Oklahoma State University, Stillwater, OK-74078}
\date{\today}
\begin{abstract}
Using resonant interaction between atoms and the field in a high
quality cavity, we show how to generate a superposition of many
mesoscopic states of the field. We study the quasi-probability
distributions and demonstrate the nonclassicality of the
superposition in terms of the zeroes of the Q-function as well as
the negativity of the Wigner function. We discuss the decoherence
of the generated superposition state. We propose homodyne
techniques of the type developed by Auffeves et al [Phys. Rev.
Lett. 91, 230405 (2003)] to monitor the superposition of many
mesoscopic states.
\end{abstract}
 \maketitle
\section{Introduction}
The interaction of a single atom with a high quality cavity has
yielded many important results which can be understood in terms of
the Jaynes-Cummings model \cite{jcm}. The advances in this field
are extensively reviewed in the literature
\cite{eberly,haroche-rmp,walther,berman}. The generation of a
superposition of mesoscopic coherent states has a fundamental
place in quantum theory as such a state exhibits quantum
interferences and the nonclassical character of the radiation
field \cite{SCHLEICH,KNIGHT}. Eiselt and Risken \cite{RISKEN} had
discovered that if a cavity contains a coherent field with large
photon numbers, say of the order of $10$, then the state of the
field for certain times splits into two parts. Each part can be
characterized approximately by a coherent state. Several authors
have studied many aspects of such splittings
\cite{GEABANACLOCHE,auffeves}. Auffeves et al \cite{auffeves} made
a successful observation of this splitting. They also devised a
novel homodyne method to observe interferences. We note that
previously such superpositions were produced using dispersive
interactions in a high quality cavity \cite{GERRY,DAVIDOVICH} or
by using Raman transitions between the center of mass degrees of
freedom of trapped ions \cite{WINELAND}.

Earlier studies of the superpositions of more than two coherent
states have found many novel features of such states. For example,
Zurek \cite{ZUREK} noticed that  such superpositions lead to
structures in phase space which are smaller than Planck's
constant. Clearly, we need efficient methods to produce such
superpositions. One of the early suggestions \cite{TARA} for the
production of such states was through the passage of a field in a
coherent state through  a Kerr medium. However Kerr nonlinearities
are usually too small. Another possibility is via the dispersive
interaction \cite{our,GERRY} in the cavity. In this paper we
present yet another possibility by using the resonant interaction
between the atom and the cavity. We show how successive passage of
atoms can be used to produce superpositions involving many
coherent states. We specifically concentrate on a superposition of
four coherent states.

The organization of the paper is as follows. In Sec II we present
the details of our proposal to produce a superposition of four
coherent states. We examines the Wigner function and the
Q-function for such states. We present a comparison of exact and
approximate phase space distribution functions. We further study
zeroes of the Q-function which are a signature of the nonclassical
properties of the field. In Sec. III we show how the passage of
the third atom can be used to monitor the superposition of four
coherent states. In Sec.IV we examine the scale over which such a
superposition can decohere.
\section{Preparation of a superposition of four  mesoscopic states of the field}
In a recent experiment, Auffeves {\it et al} \cite{auffeves} have
observed a superposition of two distinguishable states of the
field in a high quality cavity using resonant interaction between
an atom and the field inside the cavity. This observation is in
agreement with the theoretical prediction of Eiselt and Risken
\cite{RISKEN}. When a two level Rydberg atom interacts with a
microwave field, it splits the field into two parts whose phases
move in opposite directions. If the interaction time is chosen
such that the phase difference between the split parts becomes
$\pi$, then the cavity field can be projected into a superposition
similar to a cat state, $|\alpha\rangle+|-\alpha\rangle$.

In this section, we show that  the above method can be used for
the preparation of a superposition of four mesoscopic states of
the field. We consider a two level Rydberg atom having its higher
energy state $|e\rangle$ and lower energy state $|g\rangle$ and
the cavity has a strong coherent field $|\alpha\rangle$. The atom
passes through the cavity and interacts resonantly with the field.
The Hamiltonian for the system in the interaction picture is
written as
\begin{equation}
H=\hbar g\left(|e\rangle\langle g| a+a^{\dag}|g\rangle\langle
e|\right), \label{ham}
\end{equation}
where $g$ is the coupling constant for the atom with the cavity
field, and $a(a^{\dag})$ is the annihilation (creation) operator.
The state of the atom-cavity system is written as
\begin{equation}
|\psi(t)\rangle=\sum_n
\left(c_{en}(t)|e,n\rangle+c_{gn}(t)|g,n\rangle\right).
\end{equation}
Using Hamiltonian (\ref{ham}), the Schrodinger equation in terms
of $c_{en}$ and $c_{gn}$ is
\begin{eqnarray}
&&\dot{c}_{en-1}=-ig\sqrt{n}c_{gn},\\
&&\dot{c}_{gn}=-ig\sqrt{n}c_{en-1}.
\end{eqnarray}
We assume that the atom enters the cavity in its lower state
$|g\rangle$ and after interacting with the field for time $t_1$,
it is detected in the same state $|g\rangle$. Thus, effectively,
the atom absorbs no photon but it projects the cavity field into
the state
\begin{eqnarray}
|\psi_c\rangle&=&\sum_n c_n\cos(g\sqrt{n}t_1)|n\rangle,\\
 &=&\frac{1}{2}\sum_n c_n e^{ig\sqrt{n}t_1}|n\rangle+c_n
 e^{-ig\sqrt{n}t_1}|n\rangle,\\
c_n&=&\frac{\alpha^n}{\sqrt{n!}}e^{-|\alpha|^2/2}\nonumber.
\end{eqnarray}
As a result the cavity field splits into two parts whose phases
move in directions opposite to each other. Now we consider the
passage of a second identical atom through the cavity. The second
atom enters the cavity in its lower state $|g\rangle$ and after
interacting with the field for time $t_2$, is detected in the same
state $|g\rangle$. The state of the field inside the cavity after
passing the second atom is
\begin{eqnarray}
\label{twoatom}
|\psi_c'\rangle&=&\sum_nc_n\cos(g\sqrt{n}t_1)\cos(g\sqrt{n}t_2)|n\rangle,\\
&=&\frac{1}{4}\sum_nc_ne^{ig\sqrt{n}(t_1+t_2)}|n\rangle+c_ne^{-ig\sqrt{n}(t_1+t_2)}|n\rangle\nonumber\\
&&+c_ne^{ig\sqrt{n}(t_1-t_2)}|n\rangle+c_ne^{-ig\sqrt{n}(t_1-t_2)}|n\rangle.
\label{part}
\end{eqnarray}
Thus after passing second atom, the state of the field inside the
cavity splits into four parts.

In the coherent state $|\alpha\rangle$, the photon distribution
follows Poisson statistics, so in Eq.(\ref{part}), most of the
contribution to the summation comes from the terms
$n\approx|\alpha|^2$. Thus we can expand $\sqrt{n}$ in phase terms
around the average number of photons $\bar{n}=|\alpha|^2$ in
Eq.(\ref{part}). In fact for $\bar{n}\sim10$, only the terms up to
second order in $(n-\bar{n})$ are significant and other terms are
negligible.
\begin{equation}
\sqrt{n}=\sqrt{\bar{n}}+\frac{n-\bar{n}}{2\sqrt{\bar{n}}}-\frac{(n-\bar{n})^2}{8\bar{n}^{3/2}}.
\label{expan}
\end{equation}
If we substitute the value of $\sqrt{n}$ from Eq.(\ref{expan}) in
Eq.(\ref{part}), the term proportional to $n$ will change the
phase of the coherent field while the second and higher order
terms in $(n-\bar{n})$ will distort the shape of the coherent
state in phase space. For simplification, in order to understand
the nature of the generated superposition state, we do not
consider the distortion in the coherent state. Then
Eq.(\ref{part}) can be approximated by
\begin{eqnarray}
\label{meso} |\psi'_c\rangle &=&\frac{1}{4}\left[
e^{i(\eta_1+\eta_2)}|\alpha e^{i(\theta_1+\theta_2)}\rangle+
e^{-i(\eta_1+\eta_2)}|\alpha e^{-i(\theta_1+\theta_2)}\rangle+\right.\nonumber\\
&&\left.e^{i(\eta_1-\eta_2)}|\alpha
e^{i(\theta_1-\theta_2)}\rangle+
e^{-i(\eta_1-\eta_2)}|\alpha e^{-i(\theta_1-\theta_2)}\rangle\right];\\
\eta_i&=&\frac{gt_i\sqrt{\bar{n}}}{2},~\theta_i=\frac{gt_i}{2\sqrt{\bar{n}}},~i=1,2
\label{eta}
\end{eqnarray}
If we choose interaction times $t_1$ and $t_2$ such that
$\theta_1=\pi/2$ and $\theta_2=\pi/4$, we get the superposition of
four mesoscopic coherent states placed in the east, west, north
and south directions in phase space.
\begin{eqnarray}
|\psi'_c\rangle
&=&\frac{1}{4}\left[e^{-i(\eta_1-\eta_2)}|\alpha'\rangle
+e^{i(\eta_1+\eta_2)}|-\alpha'\rangle+
e^{i(\eta_1-\eta_2)}|i\alpha'\rangle+\right.\nonumber\\
&&\left.e^{-i(\eta_1+\eta_2)}|-i\alpha'\rangle \right];
\label{rmeso}
\end{eqnarray}
where we set $\alpha=\alpha'e^{i\pi/4}$.

 Now we calculate the
Wigner distribution for the state (\ref{twoatom}). The Wigner
distribution for the state having density matrix $\rho$ can be
obtained using coherent states as \cite{AGARWAL}
\begin{eqnarray}
W(\gamma)=\frac{2}{\pi^{2}}e^{2|\gamma|^{2}}\int \langle
-\beta|\rho|\beta\rangle
e^{-2(\beta\gamma^{*}-\beta^{*}\gamma)}d^{2}\beta. \label{defn}
\end{eqnarray}
The density matrix $\rho_c$ for state (\ref{twoatom}) in terms of
number states is
\begin{eqnarray}
\rho_c=\sum_{n,m}\frac{\alpha^n\alpha^{*
m}}{\sqrt{n!m!}}e^{-|\alpha|^2}
\cos(gt_1\sqrt{n})\cos(gt_2\sqrt{n})\nonumber\\
\cos(gt_1\sqrt{m})\cos(gt_2\sqrt{m})|n\rangle\langle m|.
\label{rho}
\end{eqnarray}
Using equations (\ref{defn}) and (\ref{rho}), the Wigner
distribution for the state (\ref{twoatom}) is
\begin{eqnarray}
&&W(\gamma)=\frac{2e^{2|\gamma|^2}}{\pi^2}\sum_{n,m}\frac{\alpha^n\alpha^{*
m}}{n!m!}e^{-|\alpha|^2}
\cos(gt_1\sqrt{n})\nonumber\\
&&\cos(gt_2\sqrt{n})\cos(gt_1\sqrt{m})\cos(gt_2\sqrt{m})\nonumber\\
&&\int(-\beta^*)^n\beta^m
e^{-|\beta|^2}\exp[-2(\beta\gamma^*-\beta^*\gamma)]d^2\beta.
\label{big}
\end{eqnarray}
After evaluating the integral, Eq.(\ref{big}) is simplified to the
form
\begin{eqnarray}
&&W(\gamma)=\frac{2e^{2|\gamma|^2}}{\pi}\sum_{n,m}
\frac{(-1)^{n+m}\alpha^n\alpha^{*m}}{2^{n+m}n!m!}e^{-|\alpha|^2}
\cos(gt_1\sqrt{n})\nonumber\\
&&\cos(gt_2\sqrt{n})\cos(gt_1\sqrt{m})\cos(gt_2\sqrt{m})
\frac{\partial^{n+m}}{\partial\gamma^n\partial\gamma^{*m}}e^{-4|\gamma|^2}.
\label{rwigner}
\end{eqnarray}
In the Fig.\ref{fig1} we show the Wigner distributions for the
generated superposition state (\ref{part}) as well as for the
approximated state (\ref{meso}) using some typical values of
parameters. There are four patches at the corners corresponding to
four mesoscopic states of the field and between each pair of
states of the field there are interference fringes indicating the
coherence between the states. In the central part there are
subplanck structures as noticed by Zurek \cite{ZUREK} which form
as a result of quantum interference between the two diagonal
pairs. The comparison of Fig.\ref{fig1} (a) and (b) shows that a
significant squeezing perpendicular to the arc of the circle
$|z|=|\alpha|$ occurs due to the effects of the higher order terms
in $(n-\bar{n})$ (see Eq. (\ref{expan})). Squeezing in the
resonant Jaynes-Cummings model \cite{squeezed} has been studied
very well earlier. As a result of small differences in the field
statistics, there are differences in the interference patterns. In
Fig.\ref{fig2}, the Q-distributions for the states (\ref{twoatom})
and (\ref{meso}) are shown with the same parameters used in
Fig.\ref{fig1}. We select the interaction times such that there is
no overlapping between two states of the field. A comparison of
Fig.\ref{fig2} (a) and (b) shows that the states of the field
corresponding to the phases $\pm g\sqrt{n}(t_1+t_2)$ [see
(\ref{part})] in the generated mesoscopic state have more spread
along the circle $|z|=|\alpha|$ and squeezing perpendicular to it
in phase space because of larger distortion terms. Thus the split
states of the field in the generated superposition state are
situated at the same position as in the approximated state but
with changed shape.

We further mention that after passing $N$ atoms through the cavity
and properly selecting the interaction times we can generate the
superposition of $2^N$ mesoscopic states of the field placed along
the arc of a circle of radius $|\alpha|$ in phase space. In
Fig.\ref{fig3} we show the Q-distribution for the generated state
of the field after passing three atoms through the cavity. It is
clear that the generated state is a coherent superposition of
eight mesoscopic states.

The relation between the Q-distribution and the P-distribution for
a state is given by
\begin{equation}
Q(\gamma)=\int P(\alpha)e^{-|\alpha-\gamma|^2}d^2\alpha.
\label{relation}
\end{equation}
From Eq. (\ref{relation}), it is clear that for $Q=0$ the
P-distribution will oscillate between $+ve$ and $-ve$ values. The
negative value of $P$ is a signature of the nonclassical nature of
the state. Thus the exact zeros of the Q-distribution are also
signatures of nonclassical nature. Here it will be interesting to
analyze the exact zeros of the Q-distribution of the approximated
state (\ref{rmeso}). The Q-distribution for state (\ref{rmeso}) is
\begin{eqnarray}
Q(\gamma)=\frac{1}{\pi}\left|\langle \gamma|\alpha'\rangle
e^{-i(\eta_1-\eta_2)}+\langle \gamma|-\alpha'\rangle
e^{i(\eta_1+\eta_2)}\right.\nonumber\\
\left.+\langle \gamma|i\alpha'\rangle e^{i(\eta_1-\eta_2)}+\langle
\gamma|-i\alpha'\rangle e^{-i(\eta_1+\eta_2)}\right|^2.
\label{qdis}
\end{eqnarray}
The exact zeros of $Q(\gamma)$ will be given by
\begin{eqnarray}
&&\left|\langle \gamma|\alpha'\rangle
e^{-i(\eta_1-\eta_2)}+\langle \gamma|-\alpha'\rangle
e^{i(\eta_1+\eta_2)}\right.\nonumber\\
&&\left.+\langle \gamma|i\alpha'\rangle
e^{i(\eta_1-\eta_2)}+\langle \gamma|-i\alpha'\rangle
e^{-i(\eta_1+\eta_2)}\right|=0. \label{cond}
\end{eqnarray}
Thus the Q-distribution shows nonclassical behavior at all phase
points $\gamma$ satisfying the condition (\ref{cond}). For example
if we take $\alpha'$ to be real and observe the Q-distribution
along the line $\gamma=|\gamma|e^{i\pi/4}$ in phase space, the
condition for nonclassicality (\ref{cond}) simplifies to
\begin{eqnarray}
&&e^{-\frac{|\gamma|\alpha'}{\sqrt{2}}}\cos\left[\eta_1+\eta_2+\frac{|\gamma|\alpha'}{\sqrt{2}}\right]\nonumber\\
&&+e^{\frac{|\gamma|\alpha'}{\sqrt{2}}}\cos\left[\eta_1-\eta_2+\frac{|\gamma|\alpha'}{\sqrt{2}}\right]=0.
\label{scond}
\end{eqnarray}
Now using the values of $\eta_1=\pi|\alpha'|^2/2$,
$\eta_2=\pi|\alpha'|^2/4$ (see Eq.(\ref{eta})), the condition
(\ref{scond}) can be rewritten as the simultaneous equations
\begin{eqnarray}
&&\frac{|\gamma|}{\sqrt{2}\alpha'}+\frac{3\pi}{4}=(2n_1+1)\frac{\pi}{2\alpha'^2},\nonumber\\
&&\frac{|\gamma|}{\sqrt{2}\alpha'}+\frac{\pi}{4}=(2n_2+1)\frac{\pi}{2\alpha'^2};~~n_i=1,2,..
\label{equation}
\end{eqnarray}
The solution of the equations (\ref{equation}) gives
$\alpha'^2=2(n_1-n_2)$ thus $\alpha'^2$ must be an even integer
and the values of $|\gamma|$ are given by
\begin{equation}
|\gamma|=\frac{\pi}{2\sqrt{(n_1-n_2)}}(3n_2-n_1+1);~~n_1>n_2.
\end{equation}
\section{Detection of the generated superposition of mesoscopic states of the field}
In the previous section, we have shown how the cavity field can be
projected into a superposition of $2^N$ mesoscopic  states of the
field after passing $N$ atoms through the cavity. The generated
state in the cavity can be detected by the conditional
probabilities of detection of the atoms used in the preparation
itself as the cavity field is entangled with the atomic states. An
elegant method can also be homodyne detection \cite{auffeves}
which can be implemented in the same experimental set up. After
preparing the cavity in the desired superposition state, a
resonant external coherent field $|\beta\rangle$ is injected into
the cavity. For the sake of simplicity let us assume that two
atoms are passed through the cavity in the preparation of the
mesoscopic state of the field (Eq.(\ref{twoatom})). After adding
the external field, the state of the resultant field in the cavity
is
\begin{eqnarray}
|C_h\rangle&=&\sum_n
c_n\cos(gt_1\sqrt{n})\cos(gt_2\sqrt{n})D(\beta)|n\rangle,\nonumber\\
&=&\sum_m\sum_n c_n\cos(gt_1\sqrt{n})\cos(gt_2\sqrt{n})
\langle m|D(\beta)|n\rangle|m\rangle,\nonumber\\
\label{disp}
&=&\sum_m F_m|m\rangle\\
\label{fm} F_m&=&\sum_n
c_n\cos(gt_1\sqrt{n})\cos(gt_2\sqrt{n})\langle
m|D(\beta)|n\rangle,
\end{eqnarray}
where $D(\beta)\equiv e^{\beta a^{\dag}-\beta^* a}$ is
displacement operator. Now we bring the third atom in its lower
energy state $|g\rangle$ to probe the cavity field. The
probability of detecting the probe atom in its lower state
$|g\rangle$ after crossing the cavity in time $t_p$ is
\begin{equation}
P_g=\sum_m|F_m|^2\cos^{2}(gt_p\sqrt{m}).
\end{equation}
The interaction time $t_p$ for the probe atom is selected such
that if there are photons in the cavity it leaves the cavity in
its higher energy state $|e\rangle$ with larger probability. We
have shown in the earlier section that all the states of the field
in the superposition lie on a circle of radius $|\alpha|$ so if we
choose the external field $|\beta\rangle$ having amplitude
$|\alpha|$ and phase $\phi$, the probe atom will leave the cavity
in its ground state with larger probability when the value of
$\pi+\phi$ will match to the phases of the states of the field in
the generated superposition . Thus the probability of the probe
atom leaving the cavity in its lower state $|g\rangle$ would, as a
function of $\phi$, have peaks corresponding to the positions of
the centers of the superposed mesoscopic states. In
Fig.\ref{fig4}, we plot the probability of detecting the probe
atom in its lower state with $\phi$. It shows four peaks at the
positions of the four states of the field in the generated
superposition state. The small oscillations in the background are
because of the interference effects of residual field components
after adding the external field to the cavity.
\begin{figure}
\centering
\caption{(Color online) The Wigner distributions $W(\gamma)$ for
(a) the generated state (\ref{part}) and (b) the approximated
state (\ref{meso}), using parameters $\alpha=4$, $gt_1=3.7\pi$,
$gt_2=1.9\pi$.} \label{fig1}
\end{figure}

\begin{figure}
\centering
\caption{ (Color online) The Q-distribution function $Q(\gamma)$
for (a) the generated state (\ref{part}) and (b) the approximated
state (\ref{meso}), using same parameters as in Fig.1.}
\label{fig2}
\end{figure}

\begin{figure}
\centering
\caption{(Color online) The Q-distribution function $Q(\gamma)$
for the generated state after passing three atoms through the
cavity, for $\alpha=8$. The interaction times for the first atom,
second atom and the third atom are chosen such that $gt_1=8\pi$,
$gt_2=4\pi$, $gt_3=2\pi$.} \label{fig3}
\end{figure}
\begin{figure}[h]
\centering
\includegraphics[width=3in]{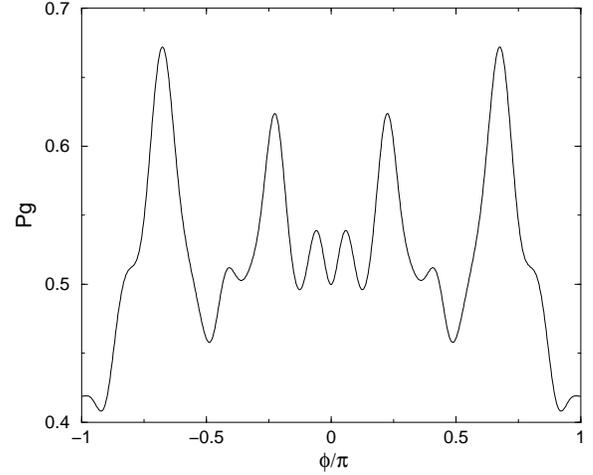}
\caption{The probability of detecting probe atom in its ground
state as a function of $\phi$ for the generated superposition
(\ref{twoatom}). The parameters used are same as in Fig.\ref{fig1}
and the interaction time for the probe atom is selected such that
$gt_p=1.5 \pi$.} \label{fig4}
\end{figure}
\begin{figure}[b]
\centering
\caption{The decoherence of the approximated state (\ref{rmeso})
in terms of Wigner function at different times, (a) for $\kappa
t=0$, (b) for $\kappa t=1/2|\alpha|^2$, (c) for $\kappa
t=1/|\alpha|^2$, (d) for $\kappa t =2/|\alpha|^2$, for
$|\alpha|=4$.} \label{fig5}
\end{figure}
\section{decoherence of the generated superposition state}
Next we study the decoherence of the generated superposition state
(\ref{part}). We are interested in the coherent superposition of
four well separated mesoscopic states of the field. The
decoherence of such a state will be equivalent to the decoherence
of the state (\ref{rmeso}). This can be done using the master
equation
\begin{equation}
\dot{\rho}=-\frac{\kappa}{2}(a^{\dag}a\rho-2a\rho a^{\dag}+\rho
a^{\dag}a),
\end{equation}
where $\kappa$ is cavity field decay parameter and we carry
analysis in the absence of thermal photons. For initial state
(\ref{rmeso}) we find the density matrix after time $t$
\begin{widetext}
\begin{eqnarray}
\label{decoh}
\rho(t)&=&\frac{1}{16}\left[\left(|\alpha_t\rangle\langle\alpha_t|+|-\alpha_t\rangle\langle-\alpha_t|
+|i\alpha_t\rangle\langle
i\alpha_t|+|-i\alpha_t\rangle\langle-i\alpha_t|\right)\right.\nonumber\\
&+&\left.e^{-2|\alpha|^2(1-e^{-\kappa
t})}\left(|\alpha_t\rangle\langle-\alpha_t|e^{-2i\eta_1}+|-\alpha_t\rangle\langle\alpha_t|e^{2i\eta_1}+
|i\alpha_t\rangle\langle-i\alpha_t|e^{2i\eta_1}
+|-i\alpha_t\rangle\langle i\alpha_t|e^{-2i\eta_1}\right)\right.\nonumber\\
&+&\left.e^{-|\alpha|^2(1-i)(1-e^{-\kappa
t})}\left(|\alpha_t\rangle\langle
i\alpha_t|e^{-2i(\eta_1-\eta_2)}+
|-i\alpha_t\rangle\langle\alpha_t|e^{-2i\eta_2}+|-\alpha_t\rangle\langle-i\alpha_t|e^{2i(\eta_1+\eta_2)}
+|i\alpha_t\rangle\langle-\alpha_t|e^{-2i\eta_2}\right)\right.\nonumber\\
&+&\left.e^{-|\alpha|^2(1+i)(1-e^{-\kappa
t})}\left(|i\alpha_t\rangle\langle\alpha_t|e^{2i(\eta_1-\eta_2)}+|\alpha_t\rangle\langle-i\alpha_t|e^{2i\eta_2}
+|-i\alpha_t\rangle\langle-\alpha_t|e^{-2i(\eta_1+\eta_2)}+|-\alpha_t\rangle\langle
i\alpha_t|e^{2i\eta_2}\right)\right];\\
\alpha_t&=&\alpha' e^{-\kappa t/2}.\nonumber
\end{eqnarray}
\end{widetext}
In Eq.(\ref{decoh}) the second, third and the fourth terms reflect
the coherent character of the superposition. These are the terms
which decohere due to interaction with the environment. The
contribution to the Wigner function from the second term in
Eq.(\ref{decoh}) is
\begin{eqnarray}
\frac{e^{-2|\gamma|^2-2|\alpha|^2(1-e^{-\kappa
t})}}{4\pi}\left\{\cos[\eta_1+i(\alpha' \gamma^*-\alpha'^*
\gamma)]\right.\nonumber\\
\left.+\cos[\eta_1+(\alpha' \gamma^*+\alpha'^* \gamma)]\right\},
\end{eqnarray}
 which decays as $e^{-2|\alpha|^2(1-e^{-\kappa t})}$
$\approx e^{-2|\alpha|^2\kappa t}$ for $\kappa t<<1$. This term
arises from the coherence between the pair $|\alpha'\rangle$,
$|-\alpha'\rangle$ and the pair $|i\alpha'\rangle$,
$|-i\alpha'\rangle$, and is responsible for the central sub-Planck
structures. The term in curly bracket can be written as
$\{\cos[\eta_1+2\alpha'|\gamma|\sin{\theta}]+\cos[\eta_1+2\alpha'
|\gamma|\cos{\theta}]\}$. Thus in any direction $\theta\ne n\pi/2$
one has an interference pattern which arises from two cosine terms
with different periodicity. Thus the sub-Planck structures
decohere as $e^{-2|\alpha|^2\kappa t}$. The third and the fourth
terms in Eq.(\ref{decoh}) show the coherence between other pairs
of coherent states, and decay as $e^{-|\alpha|^2\kappa t}$. In
Fig.\ref{fig5} we plot the decoherence of the approximated state
(\ref{rmeso}) in terms of the Wigner function at different times.
As time progresses in Fig.\ref{fig5} from (a) to (d), the central
interference patterns decay faster and disappear earlier than the
interference fringes between the coherent states, say
$|\alpha\rangle$ and $|i\alpha\rangle$, disappear. This is clear
from the equation (\ref{decoh}) that the central interference
patterns decohere two times faster than the interference fringes
between the coherent states like $|\alpha\rangle$ and
$|i\alpha\rangle$.

\section{conclusions}
In this paper we have shown the possibility for generating the
superposition of four mesoscopic states of the field using
resonant interaction between atoms and the field in a cavity. We
have discussed the properties of the quasi-probability
distributions of the generated state and compared with the
superposition of four coherent states. We have discussed the time
scale over which the state decoheres  and shown that the generated
state can be monitored using homodyne detection techniques.
Another way to detect such superposition is by doing tomography
\cite{KIM} of such states.

\end{document}